\documentclass[submission,copyright,creativecommons]{eptcs}
\providecommand{\event}{LSFA 2026} 

\usepackage{iftex}

\ifpdf
\usepackage{underscore}         
\usepackage[T1]{fontenc}        
\else
\usepackage{breakurl}           
\fi
\usepackage[utf8]{inputenc}
\usepackage[T1]{fontenc}
\usepackage{amsmath,amssymb}
\usepackage[normalem]{ulem}
\usepackage{bussproofs}
\usepackage{comment}
\usepackage{amsmath, soul}
\usepackage{amsthm}
\usepackage{amsfonts}
\usepackage{amssymb}
\usepackage{mathpartir}

\RequirePackage{color}
\usepackage[all]{xy}

\newcommand{\newor}{\,\tilde{\lor}\,}

\newtheorem{theorem}{Theorem}
\newtheorem{definition}{Definition}
\newtheorem{corollary}{Corollary}
\newtheorem{remark}{Remark}
\newtheorem{lemma}{Lemma}

\title{A Logical 3-valued Semantics for Nondeterministic Choice}

\author{Alessandro Aldini
\institute{Dipartimento di Scienze Pure e Applicate\\
Universit\`a di Urbino\\
Urbino, Italy}
\email{alessandro.aldini@uniurb.it}
\and
Pierluigi Graziani 
\institute{Dipartimento di Scienze Pure e Applicate\\
Universit\`a di Urbino\\
Urbino, Italy}
\email{pierluigi.graziani@uniurb.it}
\and
Claudio Antares Mezzina 
\institute{Dipartimento di Scienze Pure e Applicate\\
Universit\`a di Urbino\\
Urbino, Italy}
\email{claudio.mezzina@uniurb.it}
\and
Gandolfo Vergottini
\institute{Dipartimento di Scienze Pure e Applicate\\
Universit\`a di Urbino\\
Urbino, Italy}
\email{gandolfo.vergottini@uniurb.it}
}

\def\titlerunning{A Logical 3-valued Semantics for Nondeterministic Choice}
\def\authorrunning{A. Aldini, P. Graziani, C.A. Mezzina, G. Vergottini}
\begin{document}

	\maketitle
	
\begin{abstract}
We propose a logical formalisation of computational errors in reactive, nondeterministic systems. To this aim, we introduce a new three-valued symmetric nondeterministic disjunction, designed to provide a faithful logical representation of the nondeterministic choice arising in concurrent computations. The connective is defined within the framework of nondeterministic matrices (Nmatrices) and derives from a minimal combination of Kleene’s tolerant semantics and Bochvar’s symmetric error persistence, thereby eliminating the residual asymmetry induced by sequential evaluation strategies such as McCarthy’s logic. The resulting semantics admits genuinely nondeterministic outcomes in mixed cases involving errors, while preserving commutativity and operational symmetry.
\end{abstract}

\section{Introduction}

Three-valued logics have long been used to model partiality, undefinedness, and, in particular, failure in computational systems. Kleene’s strong three-valued logic~\cite{Fitting94, Kleene} interprets the third truth value as genuine indeterminacy, compatible with parallel or independent computation.
Bochvar’s internal logic~\cite{Bochvar, Finn93} treats the third value as \emph{infectious}, meaning that if it occurs in any subformula, it propagates to the whole expression, thereby modelling situations in which any local failure collapses the entire computation. McCarthy’s sequential logic~\cite{Konikowska96,McCarthy} captures the behaviour of short-circuit evaluation in sequential programming languages, where an error may halt evaluation depending on its position. Each of these systems provides a coherent account of error propagation, but each does so under a fundamentally deterministic interpretation of logical connectives, in which every connective is interpreted as a total function mapping each combination of input values to a unique output, rather than as a relation allowing multiple possible outcomes. This assumption is partial and restrictive when these logical paradigms are applied to reactive systems, which are characterized by concurrent computations and nondeterminism, according to which different computational branches may evolve independently, interact, or fail locally without enforcing a global system collapse. 
Hence, it is worth studying semantic frameworks where errors are treated as local phenomena and interact in a non-trivial way with nondeterministic choice.

In concurrency theory, this issue has been explicitly addressed in frameworks combining process algebra and three-valued logics. For instance, in an extension of the Algebra of Communicating Processes (ACP)~\cite{BP98}, an explicit error constant is introduced, governed by axioms expressing the fact that a local error suppresses both continuation and non-determinism, thus propagating globally through the system. From a logical standpoint, this corresponds to Bochvar's infectious interpretation of error. While adequate for centralized or monolithic systems, this approach becomes problematic in distributed settings, where local faults should not necessarily compromise all alternative behaviours. Later work in the CCS framework~\cite{AM25} rejects infectious error propagation and proposes an approach preserving admissible continuations, even in the presence of local failures. However, in this approach, standard deterministic connectives, including those of three-valued logics, are not sufficient to reflect all the properties related to nondeterministic behaviours.

In this paper, we address this issue by introducing a new nondeterministic disjunction, denoted by $\newor$, specifically designed to provide a logical formalisation of the nondeterministic choice between concurrent systems. The connective $\newor$ is commutative and admits genuinely nondeterministic evaluations in mixed cases involving error values.
A step supporting nondeterministic connectives was taken in~\cite{Avron2, Avron3} through the introduction of nondeterministic matrices (Nmatrices), which provide a uniform semantic framework for finite-valued logics with intrinsic non-determinism, further developing
an idea that was already present in implicit form in~\cite{Batens99, Crawford98}. In particular, Avron and Konikowska~\cite{Avron} showed how Kleene’s and McCarthy’s logics can be combined within a nondeterministic semantics. 
However, the non-determinism obtained in this way remains asymmetric. It reflects the left-to-right evaluation order characteristic of McCarthy’s logic and therefore fails to capture the genuinely symmetric behaviour of nondeterministic choice. In practice, the occurrence of an error in one branch should not suppress alternative continuations merely because of an imposed evaluation direction.
Our construction is grounded in the theory of nondeterministic matrices and builds upon the framework of~\cite{Avron}. The key idea is to combine Kleene’s tolerant behaviour with Bochvar’s symmetric error persistence to obtain a symmetric form of non-determinism, while avoiding the directional bias induced by McCarthy’s sequential semantics.

The contribution of this work is therefore twofold. 
From a logical perspective, we provide a systematic construction of a nondeterministic connective arising from the interaction between Kleene’s and Bochvar’s disjunctions within the framework of Nmatrices. Such an integration is sufficient to eliminate directional bias while preserving controlled non-determinism.
From a computational perspective, we show how this connective yields a faithful formalisation of nondeterministic choice, thus establishing a precise bridge between many-valued nondeterministic semantics and the algebraic theory of concurrent processes. 

The paper is organised as follows. Section~\ref{sc:prlmn} recalls the relevant three-valued logics and the theory of nondeterministic matrices. Section~\ref{sc:3vld} introduces the symmetric nondeterministic disjunction and its three-valued semantics. Section~\ref{sc:seqcalc} develops a sound and complete sequent calculus for the dynamic and static interpretations of this connective. Section~\ref{sc:fvsmt} refines the system above through a five-valued, purely deterministic system and establishes its soundness and completeness. 
Section~\ref{sc:conc} concludes the paper.

\section{Preliminaries}\label{sc:prlmn}

In this section, we briefly review the three-valued logics of Kleene, Bochvar, and McCarthy, together with the framework of nondeterministic matrices, which provides the semantic machinery needed to express non-determinism at the level of logical connectives.

\subsection{Kleene (K3), Bochvar (B3), and McCarthy (McC) logics}

Kleene introduced his three-valued semantics in the context of recursive function theory, where the third truth value $k$ represents an \emph{undefined} or \emph{undetermined} outcome of a partial computation~\cite{Kleene}. Unlike probabilistic or epistemic interpretations, Kleene’s reading of $k$ reflects a genuine lack of information: a truth value that has not yet been
determined -- see $\neg_{\mathbf{K}}$ and $\lor_{\mathbf{K}}$ in Table~\ref{tab:logics}.

In particular, Kleene’s disjunction is commutative and well-suited to modelling parallel or independent computations, where an undefined result in one branch does not block the overall evaluation.

In Bochvar’s logic, the third truth value $k$ is interpreted as \emph{meaningless} or \emph{infectious}~\cite{Bochvar}. Its semantic behaviour is maximally cautious: whenever a formula contains an occurrence of $k$ in any non-classical position, the evaluation collapses to $k$ -- see $\neg_{\mathbf{B}}$ and $\lor_{\mathbf{B}}$ in Table~\ref{tab:logics}.

In particular, Bochvar’s disjunction is commutative and enforces a \emph{fully persistent} and \emph{symmetric} propagation of the meaningless value, thus modelling a centralised form of error handling. 
McCarthy introduced his three-valued logic to model the behaviour of \emph{partial predicates} in sequential computation, with particular attention to \emph{runtime errors}~\cite{McCarthy}. The third value $k$ represents a \emph{critical} failure: a computation that cannot be completed and that may prevent the evaluation of subsequent expressions. The semantics follows a strict \emph{left-to-right} and \emph{short-circuit} evaluation strategy, characteristic of many programming languages (such as Algol-W, Ada, Haskell, and OCaml), where the second argument of a connective is evaluated only if needed.
Hence, for disjunction, 
if an error is encountered in the first argument, evaluation halts and the entire expression collapses to $k$; if the first argument already determines the outcome (e.g., is~$1$), the second argument is not evaluated.  
As a consequence, McCarthy’s disjunction is \emph{non-commutative} and captures the semantics of lazy, sequential computation typical of short-circuit operators in programming practice~\cite{Guzman90} -- see $\neg_{\mathbf{McC}}$ and $\lor_{\mathbf{McC}}$ in Table~\ref{tab:logics}.

\begin{table}[t]
\[
\begin{array}{c|c}
		x & \neg_{\mathbf{K}} x\\\hline
		0 & 1\\
		1 & 0\\
		k & k
\end{array}
\quad
\begin{array}{c|ccc}
		\lor_{\mathbf{K}} & 0 & 1 & k\\\hline
		0 & 0 & 1 & k\\
		1 & 1 & 1 & 1\\
		k & k & 1 & k
	\end{array}
\qquad
\begin{array}{c|c}
		x & \neg_{\mathbf{B}} x\\\hline
		0 & 1\\
		1 & 0\\
		k & k
\end{array}
\quad
\begin{array}{c|ccc}
\lor_{\mathbf{B}} & 0 & 1 & k\\\hline
		0 & 0 & 1 & k\\
		1 & 1 & 1 & k\\
		k & k & k & k
\end{array}
\qquad
\begin{array}{c|c}
x & \neg_{\mathbf{McC}} x\\\hline
0 & 1\\
1 & 0\\
k & k
\end{array}
\quad
\begin{array}{c|ccc}
\lor_{\mathbf{McC}} & 0 & 1 & k\\\hline
0 & 0 & 1 & k\\
1 & 1 & 1 & 1\\
k & k & k & k
\end{array}
\]
\caption{Truth tables for the three-valued logics \textsf{K3}, \textsf{B3}, and \textsf{McC}.}\label{tab:logics}
\end{table}

These three logics provide distinct perspectives on the nature and propagation of the third truth value: \textsf{K3} models non-critical uncertainty, \textsf{B3} models fully infectious undefinedness, and \textsf{McC} models critical sequential errors.
As we will see, while \textsf{K3} and \textsf{B3} are central for the nondeterministic semantics proposed in Section~\ref{sc:3vld}, 

\textsf{McC} does not function, in what follows, as a building block of the nondeterministic semantics itself. Its relevance re-emerges only in the deterministic setting of Section~\ref{sc:fvsmt}, where dedicated truth values are introduced explicitly to model the different interpretations of the three logics.

\subsection{Nondeterministic matrices}\label{sbsc:nmtr}
	
Nondeterministic matrices (Nmatrices)~\cite{Avron2,Avron3,Avron, Avron2005} provide a flexible semantic framework in which logical connectives may admit multiple possible outputs for a given input. This mechanism enables the semantic treatment of logics featuring  incomplete information, nondeterministic computation, or multiple admissible ways of evaluating a connective.
	
Let $L$ be a propositional language with $O$ the set of its $n$-ary connectives ($n \ge 0$) and $W$ the set of well-formed formulas. We use $p,q,r$ to denote propositional variables,  $\varphi,\psi,\tau$ to denote arbitrary formulas, and $\Gamma, \Delta$ to denote finite sets of formulas.
	
\begin{definition}
A \emph{nondeterministic matrix} (Nmatrix) for $L$ is a triple $\mathbf{M} = (V, D, O)$, where
$V$ is a non-empty set of truth values,
$D \subsetneq V$ is a non-empty set of designated values, and
$O$ includes, for each $n$-ary connective $\diamond$, an operation
$e_\diamond : V^n \to 2^V \setminus \{\emptyset\}$ assigning to every $n$-tuple
$(t_1,\dots,t_n)$ a non-empty set of possible outputs.
\end{definition}
	
Intuitively, $e_\diamond(t_1,\dots,t_n)$ represents all admissible semantic outcomes of applying $\diamond$ to arguments with truth values $t_1,\dots,t_n$. When each $e_\diamond$ always returns a singleton, the structure is simply a standard deterministic matrix.

\begin{remark}
Ordinary (deterministic) matrices arise as a special case in which each operation $e_\diamond$ is single-valued, i.e., it always returns singleton sets. In this situation, each $e_\diamond$ can be identified with a function $e_\diamond : V^n \to V$. Consequently the semantics becomes fully deterministic.
\end{remark}
\begin{definition}
Let $\mathbf{M} = (V, D, O)$ be an Nmatrix.
\begin{enumerate}
\item A \emph{dynamic valuation} in $\mathbf{M}$ is a function $v : W \to V$ such that for every $n$-ary connective $\diamond$ and every $\psi_1,\dots,\psi_n \in W$:
\[
v(\diamond(\psi_1,\dots,\psi_n)) \;\in\; e_\diamond\bigl(v(\psi_1),\dots,v(\psi_n)\bigr).
\tag{S}
\]
\item A \emph{static valuation} in $\mathbf{M}$ is a dynamic valuation satisfying also the following compositionality condition: for all $\psi_1,\dots,\psi_n$ and $\varphi_1,\dots,\varphi_n$, if $v(\psi_i)=v(\varphi_i)$ for all $i$, then
\[
v(\diamond(\psi_1,\dots,\psi_n)) = v(\diamond(\varphi_1,\dots,\varphi_n)).
\tag{C}
\]
\end{enumerate}
\end{definition}
	
Dynamic semantics reflects the highest degree of non-determinism: each evaluation of a connective may pick \emph{any} value allowed by $e_\diamond$ independently of previous choices. In contrast, static valuations enforce global coherence: Condition (C) ensures that the interpretation of each connective behaves like a (single-valued) function $f^\mathbf{v}_\diamond : V^n \to V$ chosen
\emph{once and for all} before evaluation begins, subject to
$f^\mathbf{v}_\diamond(t_1,\dots,t_n) \;\in\; e_\diamond(t_1,\dots,t_n)$.
	
\begin{definition}
Given an Nmatrix $\mathbf{M}$:
\begin{itemize}
\item a valuation $v$ \emph{satisfies} a formula $\psi$ (written $v \models \psi$) if $v(\psi)\in D$;
\item $v$ is a \emph{model} of a set $\Gamma$ if $v\models \gamma$ for all $\gamma\in \Gamma$.
\end{itemize}
\end{definition}
	
\begin{definition}
Let $\Gamma,\Delta$ be finite sets of formulas.
\begin{itemize}
\item A valuation $v$ \emph{satisfies} the sequent $\Gamma \Rightarrow \Delta$ (written $v \models \Gamma \Rightarrow \Delta$) if either $v\not\models\gamma$ for some $\gamma\in\Gamma$, or $v\models\delta$ for some $\delta\in\Delta$.
\item A sequent is \emph{dynamically valid} in $\mathbf{M}$, written
$\models^\mathrm{d}_{\mathbf{M}} \Gamma \Rightarrow \Delta$, if it is satisfied by all dynamic valuations in $\mathbf{M}$.
\item A sequent is \emph{statically valid} in $\mathbf{M}$ if it is satisfied by all static valuations, written $\models^\mathrm{s}_{\mathbf{M}} \Gamma \Rightarrow \Delta$.
\end{itemize}
\end{definition}
	
Dynamic and static semantics thus induce two consequence relations, $\vdash^\mathrm{d}_{\mathbf{M}}$ and $\vdash^\mathrm{s}_{\mathbf{M}}$, corresponding to local and global resolution of non-determinism, respectively.
Deterministic matrices appear as the degenerate case in which the two notions coincide.

\section{Yet another semantics for 3-valued disjunction}\label{sc:3vld}

The semantic machinery introduced so far provides a general setting for comparing different three-valued treatments of error. At this stage, however, we need to clarify the extent to which existing approaches are able to account for the behaviour of nondeterministic choice.

The most significant combination of non-determinism and three-value semantics is due to Avron and Konikowska~\cite{Avron}, who develop a systematic logical framework for reasoning about computation errors in program verification. 

The starting point of their work is the integration within a single logical framework of \textsf{McC} and \textsf{K3}.
The former provides the semantics for lazy, sequential computation, where an error immediately halts the entire process (\textit{critical} error). The latter captures the behaviour of parallel computations, where an error in one branch does not necessarily compromise the whole process (\textit{non-critical} error). 
Hence, the resulting system is capable of flexibly representing different computational strategies (lazy versus parallel), while preserving both a finite semantics and a two-sided sequent calculus of classical form.
To achieve this, the authors define a three-valued Nmatrix obtained by merging the truth tables of 
\textsf{McC} and \textsf{K3}. 
Under a dynamic semantics, the interpretation of connectives can vary locally, allowing the computational process to adopt either the Kleene or the McCarthy strategy on a case-by-case basis.
Under a static semantics, by contrast, the chosen strategy is fixed globally across all evaluations, though it remains unspecified which one is applied.
Then, the authors extend this setting to a four-valued deterministic matrix, which introduces explicitly two distinct error values ($e$, representing a critical, McCarthy-type error, and
$u$, representing a non-critical, Kleene-type error).
This richer framework makes it possible to model both types of computational failure within a single deterministic semantics, thereby unifying reasoning about different modes of error propagation in computation.

The form of non-determinism encoded in~\cite{Avron} is used to model global and local errors in the same framework. More in general, the presence of non-determinism intrinsically characterizes the behaviour of concurrent systems.

In the classical setting of structural operational semantics (sos)~\cite{Plo2004,Niel2}, 
the program defined as $(\mathsf{if}~\varphi~\mathsf{then}~S_1)~\mathsf{or}~S_2$ non-deterministically chooses between the execution of the \textsf{if}-statement and the execution of $S_2$. In this example, it is reasonable to ask what would happen if the evaluation of the Boolean formula $\varphi$ caused an error. 
Different answers are proposed in various approaches to concurrency theory, especially in the formal framework of  process description languages (\textit{process algebra}).\footnote{We recall that typical process calculi, like ACP and CCS~\cite{baeten2005brief}, include the $\mathsf{or}$-statement as an operator $P_1 + P_2$ modeling the nondeterministic choice between process $P_1$ and process $P_2$. Informally, the sos semantics of $+$ states that if $P_i$, $i = 1,2$, can make a step evolving to $P_i'$, then $P_1 + P_2$ can make the same step evolving to $P_i'$.}

In a seminal work, Bergstra and Ponse~\cite{BP98} investigate the $\mathsf{if}$-statement with three-valued semantics in the setting of the process algebra ACP, which is extended with a process term $\mu$ expressing the process terminated with an error, and a statement $\varphi :\rightarrow P$ expressing the command $\mathsf{if}~\varphi~\mathsf{then}~P$. 
On the one hand, the property 
$\mu + P = \mu$ holds (\textit{infectious error}) to state that a local error infects the overall system, thus suppressing non-determinism. On the other hand, the semantics of $\varphi$ is given in terms of a three-valued logic (here we continue to use $k$ to denote the third truth value), such that the following properties hold:
\begin{align}
	k :\rightarrow P = \mu 
    \label{axiom:mu} \\
	\mathsf{\varphi}  :\rightarrow (\mathsf{\psi}:\rightarrow P) = 
    (\mathsf{\varphi \land_{\mathbf{McC}} \psi}): \rightarrow P 
    \label{axiom:and} \\
	\mathsf{\varphi}:\rightarrow P + \mathsf{\psi}:\rightarrow P = 
    (\mathsf{\varphi \lor_{\mathbf{B}} \psi}): \rightarrow P 
    \label{axiom:or}
\end{align}
In particular, (\ref{axiom:mu}) states the relation between the third value $k$ and the error process.
In (\ref{axiom:and}), the use of the conjunction operator $\land_{\mathbf{McC}}$, which is the De Morgan dual of $\lor_{\mathbf{McC}}$, is compliant with the lazy interpretation of nested \textsf{if}-statements. If the outer condition is false, the inner condition is not evaluated, as expressed by $0 \land_{\mathbf{McC}} k = 0$. In (\ref{axiom:or}), using Bochvar's disjunction makes it clear that the intended interpretation of the error is global, that is, the occurrence of an error propagates to the whole system, as expressed by the truth table of $\lor_{\mathbf{B}}$.

In~\cite{AM25}, the process algebra CCS is extended along the same line of~\cite{BP98}, with the difference that the infectious error property is refused. In this alternative setting, (\ref{axiom:and}) holds as it is, while (\ref{axiom:or}) does not hold anymore. To recover this property, the authors' claim is that a new disjunction operator shall replace $\lor_{\mathbf{B}}$:
\begin{equation}
    \mathsf{\varphi}:\rightarrow P +
	\mathsf{\psi}:\rightarrow P = (\mathsf{\varphi \newor \psi}): \rightarrow P
    \label{axiom:newor}
\end{equation}
where $\newor$ must reflect the nondeterministic behavior of $+$, i.e., $k \newor 1$ non-deterministically returns either $k$ or 1, in the same way $k :\rightarrow P + 1 :\rightarrow P$ may either fail or continue as $P$, according to a local interpretation of the error occurrence. Obviously, since $+$ is commutative, also $\newor$ so is.
	
One may wonder whether a candidate for this new disjunction operator can be found in~\cite{Avron}, where, however, the non-determinism encoded by the dynamic three-valued matrix results from the \emph{combination} of the Kleene and McCarthy disjunctions. 
Hence, the outcome is an \emph{asymmetric} form of non-determinism, which is adequate for sequential computations, but it fails to capture the behaviour of concurrent and distributed systems.
This limitation is therefore not technical but structural. 
	
For these reasons, we introduce a genuinely \emph{symmetric} form of non-determinism, directly linked to the \emph{choice behaviour} of the nondeterministic operator---such as the process-algebraic operator $+$. 
To this end, we combine the approach developed in~\cite{Avron} with the disjunction of~\cite{Bochvar}, whose semantics embodies a fully \emph{infectious error behaviour}. The resulting operation, denoted by $\newor$, is \emph{commutative} and its truth matrix includes two entries that are not uniquely determined, namely: 
$ k \newor 1 = 1 \newor k$.
This construction yields a nondeterministic disjunction that mirrors the operational semantics of the process-algebraic choice operator $+$, thus respecting property (\ref{axiom:newor}).
In this setting, the choice between the two continuations is made non-deterministically, but---unlike in~\cite{Avron}---no possible continuation is suppressed.

\begin{remark}
Our approach departs from~\cite{Avron}, where the dynamic Nmatrix is obtained by combining $\lor_{\mathbf{K}}$ and $\lor_{\mathbf{McC}}$. In the present framework, we extend this construction by incorporating Bochvar’s disjunction $\lor_{\mathbf{B}}$, whose infectious semantics enforces symmetry and persistence of $k$. 
As a consequence, the asymmetric behaviour induced by McCarthy’s evaluation order becomes unnecessary, and $\lor_{\mathbf{McC}}$ plays no operative role in the nondeterministic setting. The combination of $\lor_{\mathbf{K}}$ and $\lor_{\mathbf{B}}$ thus provides a minimal and sufficient basis for defining the commutative nondeterministic operator $\tilde{\lor}$.

McCarthy’s logic will re-enter the picture at a later stage, within the deterministic five-valued framework of Section~\ref{sc:fvsmt}. In fact, once non-determinism has been isolated at the three-valued level, the sequential behaviour characteristic of McCarthy’s semantics can be reintroduced in a controlled and explicit way, as a specific mode of error propagation rather than as an artefact of nondeterministic evaluation.
\end{remark}

\subsection{The 3-valued Nmatrix}
The logical foundation of the system rests on a three-valued nondeterministic matrix extending the dynamic semantics introduced in~\cite{Avron}. The aim of this modification is to retain the elegant interplay between \textsf{K3} and \textsf{McC} while eliminating the residual asymmetry that characterises their disjunction. In the framework of~\cite{Avron}, non-determinism expresses an epistemic uncertainty regarding the evaluation strategy—whether sequential or parallel—thus yielding the directional behaviour $1 \lor k = 1$ but $k \lor 1 = k$. In the present approach, the connective is reinterpreted so that non-determinism becomes symmetric and genuinely operational: both outcomes are admissible, reflecting the fact that the system can evolve along either computational branch.
		
\medskip
\noindent
\textbf{Truth structure.}
The matrix is defined over the set of truth values $V = \{0, 1, k\}$, with the set of designated values $D = \{1\}$. Negation behaves classically on the extreme values and leaves the indeterminate value fixed:
$\neg 0 = 1$, $\neg 1 = 0$, $\neg k = k$.
This mirrors the behaviour of Kleene’s, McCarthy's, and Bochvar’s systems, ensuring that $k$ acts as a fixed point of uncertainty.
	
\medskip
\noindent
\textbf{Symmetric nondeterministic disjunction.}
The binary connective $\newor$ is given by the following truth table:
\[
\begin{array}{c|ccc}
\newor & 0 & 1 & k\\\hline
0 & \{0\} & \{1\} & \{k\}\\
1 & \{1\} & \{1\} & \{1,k\}\\
k & \{k\} & \{1,k\} & \{k\}
\end{array}
\]
We denote by $\mathbf{M}^3_{BMK}$ the three-valued nondeterministic matrix defined above.

The crucial modification occurs in the cells $(1,k)$ and $(k,1)$, which are now symmetric and take the value set $\{1,k\}$. This expresses that, when one component is successful and the other undefined, the resulting computation may either succeed or remain indeterminate. The connective therefore captures a form of constructive non-determinism consistent with concurrent evaluation.

\medskip
\noindent
\textbf{Dynamic semantics.} 
A valuation $v$ assigns to each formula a value in $V$. For a compound formula $\alpha\newor\beta$ we require
$
v(\alpha\newor\beta)\in \newor(v(\alpha),v(\beta)).
$

\medskip
\noindent
\textbf{Sequent satisfaction.}
A valuation $v$ satisfies a sequent $\Gamma\Rightarrow\Delta$ (written $v\models\Gamma\Rightarrow\Delta$) iff
\[
(\exists\gamma\in\Gamma)\, v(\gamma)\notin D \quad\text{or}\quad (\exists\delta\in\Delta)\, v(\delta)\in D.
\]
Intuitively, a sequent is true under a valuation if at least one formula in the antecedent fails to be designated, or if at least one formula in the succedent is designated. This definition ensures that the sequent relation remains monotonic and preserves the classical interpretation of entailment within a nondeterministic setting.

\section{Sequent calculus for $\newor$}
\label{sc:seqcalc}

Having introduced a symmetric nondeterministic semantics for disjunction, we now turn to its proof-theoretic counterpart and develop a sequent calculus that reflects this behaviour.
In the following, we first introduce the sequent calculus corresponding to the dynamic three-valued semantics and establish its soundness. We then prove completeness via a canonical valuation construction, before extending the system to the static case and proving its metaproperties.

The sequent calculus -- called $SCd_{B}$ and shown in Table~\ref{tab:nd-sq} -- is not merely a technical companion to the semantics. Its structure reflects, at the inferential level, the same rejection of directional evaluation bias that motivates the nondeterministic connective. 
In particular, the calculus extends the structure of the dynamic system proposed in~\cite{Avron}, while adapting its right-introduction rules to capture the symmetric non-determinism of the connective $\newor$.
The result is a system that must be intended as the proof-theoretic counterpart of the nondeterministic matrix
$\mathbf{M}^3_{BMK}$ introduced in the previous section. We present the following rules using a non-consecutive numbering scheme, with intentional gaps in the sequence. This is due to the fact that the present system is introduced for the dynamic nondeterministic matrix and will later be extended to the deterministic five-valued setting by adding further rules that refine and complete the calculus. We follow the same convention as in \cite{Avron} in order to maintain consistency with the existing literature.

\begin{table}[t]

\begin {mathpar}
\inferrule*[left=a1]{}{\Gamma,\alpha \Rightarrow \Delta,\alpha} \and
\inferrule*[left=a2]{}{\Gamma,\alpha,\neg\alpha \Rightarrow \Delta}
\end{mathpar}

\begin{mathpar}
\inferrule*[left=r1]{\Gamma,\alpha \Rightarrow \Delta}{\Gamma,\neg\neg\alpha \Rightarrow \Delta} \and
\inferrule*[left=r2]{\Gamma \Rightarrow \Delta,\alpha}{\Gamma \Rightarrow \Delta,\neg\neg\alpha} \and
\end{mathpar}
\begin{mathpar}
	\inferrule*[left=r3]{\Gamma,\alpha \Rightarrow \Delta \quad \Gamma,\beta \Rightarrow \Delta}{\Gamma,\alpha\newor\beta \Rightarrow \Delta} \and
\end{mathpar}
\begin{mathpar}
    \inferrule*[left=r$4_a$]{\Gamma \Rightarrow \Delta,\alpha \quad \Gamma \Rightarrow \Delta,\neg\beta}{\Gamma \Rightarrow \Delta,\alpha\newor\beta}
    \and
    \inferrule*[left=r$4_b$]{\Gamma \Rightarrow \Delta,\alpha \qquad \Gamma \Rightarrow \Delta,\beta}{\Gamma \Rightarrow \Delta,\alpha\newor\beta}
    \and
    \inferrule*[left=r5]{\Gamma \Rightarrow \Delta,\neg\alpha \quad \Gamma \Rightarrow \Delta,\beta}{\Gamma \Rightarrow \Delta,\alpha\newor\beta}
    \and
\end{mathpar}
\begin{mathpar}
    \inferrule*[left=r8]{\Gamma,\neg\alpha \Rightarrow \Delta}{\Gamma,\neg(\alpha\newor\beta)\Rightarrow \Delta}
    \and 
\inferrule*[left=r9]{\Gamma\Rightarrow\Delta,\neg\alpha \quad \Gamma\Rightarrow\Delta,\neg\beta}{\Gamma\Rightarrow\Delta,\neg(\alpha\newor\beta)}
\and 
\inferrule*[left=r10]{\Gamma,\neg\beta \Rightarrow \Delta}{\Gamma,\neg(\alpha\newor\beta)\Rightarrow \Delta}
\end{mathpar}

    \caption{Sequent calculus $SCd_{B}$.}
    \label{tab:nd-sq}
\end{table}
%
Specifically, the difference between the system of~\cite{Avron} and $SCd_{B}$ lies essentially in the right-introduction of disjunction as follows.
In~\cite{Avron}, the asymmetric behaviour of McCarthy’s disjunction is captured by a single right rule.
Here, that rule splits into the two rules $(R4_a)$ and $(R4_b)$, reflecting the symmetric non-determinism of $\newor$. Hence, the duplication of right-introduction rules for the disjunction is a direct proof-theoretic
manifestation of the symmetry that characterizes the semantics of the disjunction connective $\newor$.
The remaining rules play a standard structural role. Axioms $(A1)$-$(A2)$ and rules $(R1)$-$(R2)$ govern identity, inconsistency, and involutive negation. Rule $(R3)$ introduces disjunction on the left, rule $(R5)$ handles right-introduction in the presence of a negated disjunct, and rules $(R8)$-$(R10)$ regulate the interaction between negation and disjunction.

The proof-theoretic adequacy of the calculus with respect to the intended semantics is ensured by the following soundness and completeness results. Their proofs follow the standard strategy for dynamic calculi based on nondeterministic matrices, in the spirit of~\cite{Avron}.

\begin{theorem}[Soundness]\label{th:snd3d}
Every rule of the calculus $SCd_B$ preserves validity with respect to the dynamic semantics determined by the three-valued Nmatrix $\mathbf{M}^3_{BMK}$.
Consequently, every sequent provable in $SCd_B$ is valid under all dynamic valuations for this matrix.
\end{theorem}

\begin{theorem}[Completeness for the dynamic semantics]\label{th:cmpl3d}
If a sequent $\Gamma \Rightarrow \Delta$ is valid in every dynamic valuation for the Nmatrix $\mathbf{M}^3_{BMK}$, then it is provable in $SCd_{B}$.
Equivalently, if $\Gamma \Rightarrow \Delta$ is not provable, then there exists a dynamic valuation $v$ such that
$v(\gamma) = 1$ for all $\gamma \in \Gamma$ and $v(\delta) \neq 1$ for all $\delta \in \Delta$.
\end{theorem}
\subsection{An Example of Derivation: Disjunctive Syllogism}
To illustrate the operational behavior of the calculus $SCd_{B}$, we consider the derivation of a version of the disjunctive syllogism. This example is particularly significant as it shows how the calculus handles the interaction between the nondeterministic disjunction and negation in the presence of designated values.

\begin{prooftree}
	\AxiomC{}
	\RightLabel{($A2$)}
	\UnaryInfC{$\neg\alpha, \alpha \Rightarrow \beta$}
	\AxiomC{}
	\RightLabel{($A1$)}
	\UnaryInfC{$\neg\alpha, \beta \Rightarrow \beta$}
	\RightLabel{($R3$)}
	\BinaryInfC{$\neg\alpha, \alpha \newor \beta \Rightarrow \beta$}
\end{prooftree}

\noindent \textbf{Commentary:} The derivation proceeds by analyzing the nondeterministic disjunction on the left side via rule $(R3)$. This rule splits the proof into two branches: 
(i) the case where $\alpha$ holds, which is immediately closed by the inconsistency axiom $(A2)$ due to the presence of $\neg\alpha$ in the antecedent, and 
(ii) the case where $\beta$ holds, which is closed by the identity axiom $(A1)$ since $\beta$ directly satisfies the conclusion. 
Crucially, if $\alpha$ were an error state (non-designated), the premise $\neg\alpha$ would prevent the left branch from succeeding unless $\beta$ itself provides the necessary truth value, reflecting the error-masking capabilities of our symmetric disjunction.
\subsection{Static semantics}\label{ssc:stc}

The semantics we have presented in the previous section is dynamic. It captures local resolution, where nondeterministic choice is resolved ``on-the-fly'' at each site of the operator. This models reactive and distributed systems where different components may manifest distinct error propagation patterns due to local environmental factors. In this section, we move to a static perspective.
The computational interpretation of the static semantics corresponds to a globally fixed resolution (though potentially unknown at design-time) of nondeterministic choice. This is representative of centralized systems or specific execution environments where the evaluation policy (e.g., how the indeterminacy of $\newor$ is resolved) is globally uniform. 

For this purpose, the static semantics we propose extends the dynamic framework by imposing a global deterministic choice for the evaluation of the connective $\newor$. In the dynamic case, each occurrence of $\alpha \newor \beta$ may select, at evaluation time, any element of $\newor(v(\alpha),v(\beta))$. The static interpretation, instead, fixes \emph{once and for all} a choice function $f_{\lor} : V^{2} \to V$ such that $f_{\lor}(x,y) \in \newor(x,y)$ for all $x,y \in V$.
Accordingly, a valuation is \emph{static} when $v(\alpha \newor \beta) = f_{\lor}(v(\alpha),v(\beta)).$

By formalizing this dynamic vs.~static duality, we provide a framework for comparative analysis. The static approach facilitates verification through global invariants (as it models a stable, deterministic implementation of the nondeterministic specification), while the dynamic semantics enables precise tracking of error propagation in loosely coupled, nondeterministic environments. Relevant computational scenarios, such as debugging or fault-tolerance analysis, are affected by this interpretation in the same way as distributed and centralized computations differ in their error-handling strategies (see, e.g., \cite{Arieli2009}).

In order to capture the static behaviour at the proof-theoretic level, the calculus is enriched with two additional rules, which generalize the single static rule introduced in~\cite{Avron} into a pair of symmetric counterparts:
\begin{mathpar}
	\inferrule*[left=S1]{\Gamma\Rightarrow \Delta, \beta \quad \Gamma, \phi \Rightarrow \Delta \quad \Gamma, \neg\phi \Rightarrow \Delta}{\Gamma, \phi \newor \psi \Rightarrow \Delta, \alpha \newor \beta}
	\and 
	\inferrule*[left=S2]{\Gamma\Rightarrow \Delta, \alpha \quad \Gamma, \phi \Rightarrow \Delta \quad \Gamma, \neg\phi \Rightarrow \Delta}{\Gamma, \psi \newor \phi \Rightarrow \Delta, \alpha \newor \beta}
\end{mathpar}
Rule (S1) (already appears in~\cite{Avron}) corresponds to the deterministic propagation guided by the first argument $\phi$, while (S2) provides its symmetric counterpart driven by $\psi$. Together, they enforce a coherent global choice compatible with the commutativity of $\newor$.

The logic induced by the static semantics of the Nmatrix captures the situation in which the computational device adopts a \emph{single, fixed evaluation strategy} for the connective $\newor$, which is applied uniformly throughout the entire computation.
While the particular strategy is not specified \emph{a priori}, once it is chosen it is used consistently at every occurrence of $\newor$.
In this sense, the static semantics models a form of \emph{global determinisation of non-determinism}.
	
More concretely, suppose that for some formula $\alpha \newor \beta$ we have $v(\alpha) = 1$ and $v(\beta) = k$, and assume that the chosen static function $f_{\lor}$ satisfies $f_{\lor}(1,k) = 1$.
Then the same selection principle is applied uniformly to all other occurrences of $\newor$ with the same input values.
In particular, if another formula $\gamma \newor \delta$ satisfies
$v(\gamma)=1$ and $v(\delta)=k$, then necessarily
$v(\gamma \newor \delta) = f_{\lor}(1,k) = 1$.
Thus, unlike the dynamic case—where each occurrence may resolve non-determinism independently—the static semantics enforces a \emph{coherent, system-wide resolution strategy}.
This corresponds to the computational scenario in which the machine always handles nondeterministic choice in the same way, even though this strategy is not externally observable.

Similarly as for the dynamic setting, even in the static approach sound and completeness results hold.
	
\begin{theorem}[Soundness of the static calculus]\label{th:snd3s}
The sequent calculus obtained by adding the static rules \emph{(S1)} and \emph{(S2)} to the dynamic system is sound with respect to the class of static valuations induced by deterministic choice functions compatible with the Nmatrix of $\newor$.
\end{theorem}
    
\begin{theorem}[Completeness of the static calculus]\label{th:cmpl3s}
The sequent calculus for $\newor$ is complete with respect to the class of static valuations induced by the three-valued nondeterministic matrix.
\end{theorem}
	
\begin{remark}
The only conceptual difference with the static system in~\cite{Avron} lies in the duplication of the deterministic rule: while in~\cite{Avron} the original connective satisfies $1\lor k = 1$ and thus requires a single orientation, the symmetric connective of the present framework demands both (S1) and (S2) to preserve completeness under the commutative interpretation of $\newor$.
\end{remark}

\subsection{Decidability and finitarity}

An important meta-theoretic property of the system concerns the behaviour of the entailment relation induced by the three-valued Nmatrix introduced above. Since this matrix is finite, we can apply a general result due to Avron and Lev~\cite{Avron3} concerning nondeterministic semantics.

In the framework of Nmatrices, different consequence relations can be induced depending on the set of designated values and the properties of the entailment. We adopt a \emph{multiple-conclusion} consequence relation $\vdash_{\mathbf{M}}$ between sets of formulas, specifically following the definition of \emph{Scott consequence relation}~\cite{Avron3}.
Scott consequence relations are based on the idea of treating logical consequence as a binary relation between sets of propositions. Under this interpretation, the conclusions represent a set of alternatives that are implied, or required, by the premises.

\begin{definition}[Scott Consequence Relation]
A consequence relation $\vdash$ on a set of formulas $W$ is a binary relation between sets of formulas $\Gamma, \Delta \subseteq W$ satisfying the following axioms for all sets of formulas $\Gamma, \Delta, \Theta$:
\begin{enumerate}
\item \textbf{Strong Reflexivity:} If $\Gamma \cap \Delta \neq \emptyset$, then $\Gamma \vdash \Delta$.
\item \textbf{Monotonicity:} If $\Gamma \vdash \Delta$ and $\Gamma \subseteq \Gamma'$, $\Delta \subseteq \Delta'$ , then $\Gamma' \vdash \Delta'$.
\item \textbf{Transitivity:} If $\Gamma \vdash \Delta, \varphi$ and $\Gamma', \varphi \vdash \Delta'$, then $\Gamma, \Gamma' \vdash \Delta, \Delta'$.
\end{enumerate}
\end{definition}
\begin{theorem}[Avron--Lev]
\label{thm:avron-finitary}
If $\mathbf{M}$ is a finite Nmatrix, then its associated consequence relation
$\vdash_{\mathbf{M}}$ (and hence also the single-conclusion version
$\vdash^{1}_{\mathbf{M}}$) is finitary. That is, for every set of formulas $T$
and every formula $\varphi$,
\[
T \vdash_{\mathbf{M}} \varphi
\quad\text{iff}\quad
T_{0} \vdash_{\mathbf{M}} \varphi
\text{ for some finite } T_{0}\subseteq T .
\]
\end{theorem}

In our setting, the dynamic semantics is defined by the finite Nmatrix $\mathbf{M}^3_{BMK}=\langle V,D,\neg,\newor\rangle$. Therefore, by Theorem~\ref{thm:avron-finitary}, the semantic consequence relation associated with $\mathbf{M}^3_{BMK}$ is finitary.  Since the calculus $SCd_{B}$ is sound and complete with respect to this semantics, we immediately obtain the following corollary.

\begin{corollary}[Finitarity of $SCd_{B}$]
\label{cor:finitary-calculus}
For every set of formulas $T$ and every formula $\varphi$,
\[
T \vdash_{SCd_{B}} \varphi
\quad\text{iff}\quad
T_{0} \vdash_{SCd_{B}} \varphi
\text{ for some finite } T_{0}\subseteq T .
\]
Hence the proof system $SCd_{B}$ is finitary.
\end{corollary}

As for standard consequence relations, the consequence relation of $SCd_{B}$ is \emph{decidable} (see also \cite[Theorem 2.8]{Avron000}). The same considerations apply to the static semantics introduced in Section~\ref{ssc:stc}.

\section{Sequent calculus for the 5-valued deterministic system}\label{sc:fvsmt}

The results established so far provide a complete logical analysis of symmetric nondeterministic choice at the three-valued level. In particular, the dynamic and static semantics of the connective $\newor$ fully capture the operational behaviour of symmetric choice without committing to any fixed evaluation strategy.
In this way, we model situations in which a computing agent is unaware of the specific evaluation strategy adopted in the presence of errors. In that setting, the indeterminacy in the interpretation of $\newor$ reflects precisely the lack of control over how anomalous values are propagated.

In the present section, by contrast, we shift to a deterministic perspective, where the machine is assumed to be able to distinguish between different kinds of informational failures and to react to them according to fixed, strategy-dependent rules. To this end, we introduce a standard five-valued matrix, denoted by $SC_{5}$. This should be read as a deterministic refinement of the nondeterministic semantics developed above. Its purpose is to make explicit—within a fully deterministic setting—the different modes of computational failure that remain implicit at the nondeterministic level.

The system $SC_{5}$ provides a uniform framework in which the behaviours associated with the third truth value in the three-valued logics of Kleene, Bochvar, and McCarthy are separated and explicitly represented. What was previously encoded by a single indeterminate value is here refined into distinct types of error, allowing us to capture not only the presence of a failure, but also its logical and computational nature.

From this perspective, $SC_{5}$ extends the four-valued deterministic framework of~\cite{Avron} by enriching it with a finer-grained taxonomy of errors, designed to capture more faithfully the dynamics of fault propagation in computational processes.
Accordingly, it is not intended as an independent logical proposal, but as a conceptual completion of the nondeterministic analysis.

In this setting, McCarthy’s disjunction regains a central role. While in the nondeterministic semantics its asymmetric behaviour is neutralised by symmetry and nondeterministic choice, here it becomes essential to model order-sensitive computations. Together with Kleene’s tolerant behaviour and Bochvar’s infectious one, McCarthy’s connective captures a genuinely sequential form of error propagation, in which the evaluation order directly affects the outcome. 

In this sense, McCarthy’s logic recovers a central conceptual role within the deterministic five-valued setting, providing the essential bridge between logical non-classicality and operational, step-by-step computation.

Formally, the semantics for our deterministic approach is based on the deterministic five-valued matrix $\mathbf{M}^5_{BMK}=(V, D, O)$, where $V=\{0,1,u,e,k\}$, $D=\{1\}$, and $O=\{\neg, \lor\}$. The values $0$ and $1$ represent the classical Boolean values (false and true), while $u$, $e$, and $k$ denote, respectively, the Kleene-, McCarthy-, and Bochvar-type error states.
Negation behaves conservatively, acting classically on the Boolean values and leaving each non-classical value fixed:
$$\neg 0 = 1, \neg 1 = 0, \neg u = u, \neg e = e, \neg k = k.$$
The non-commutative disjunction $\lor$ is defined by the following table:
\[
\begin{array}{c|ccccc}
\lor & 0 & 1 & u & e & k\\\hline
0 & 0 & 1 & u & e & k\\
1 & 1 & 1 & 1 & 1 & k\\
u & u & 1 & u & e & k\\
e & e & e & e & e & k\\
k & k & k & k & k & k
\end{array}
\]

Intuitively, the connective captures a form of sequential computation where the left-hand argument may influence how the right-hand one is evaluated, as in McCarthy’s short-circuit semantics.
However, unlike the four-valued case, the present system distinguishes three levels of failure:
\begin{itemize}
\item \textbf{Kleene errors} ($u$) represent soft inconsistencies or indeterminate outcomes that do not affect the success of an otherwise valid computation.
Hence, $1 \lor u = u \lor 1 = 1$: the system successfully recovers.
\item \textbf{McCarthy errors} ($e$) correspond to operational faults that depend on evaluation order: if the error occurs first, it propagates; otherwise, it may be bypassed. This introduces non-commutativity in the disjunction.
\item \textbf{Bochvar errors} ($k$) are critical, absorbing failures that block any further computation regardless of context: $x \lor k = k \lor x = k$ for all $x\in V$.
\end{itemize}
	
In computational terms, these values describe systems capable of distinguishing the severity and propagation conditions of errors: Kleene-type uncertainty, McCarthy-type context sensitivity, and Bochvar-type fatal inconsistency.
The resulting calculus $SC_{5}$ -- see Table~\ref{tab:sc5} -- provides a sequent-level characterization of this graded notion of failure, generalizing the deterministic logic $\mathbf{M}_{4}^{MK}$ presented in~\cite{Avron} to a setting where error management becomes explicitly ordered and semantically stratified.

\subsection{Soundness and completeness of the calculus $SC_{5}$}

The sequent calculus $SC_{5}$ largely inherits its axioms and inference rules from the three-valued dynamic system introduced in Section~\ref{sc:seqcalc}. Axioms $(A1)$-$(A2)$, the rules for negation, and the basic introduction rules for disjunction coincide with those of $SCd_{B}$.
The only genuinely new rules are $(R6)$–$(R7)$ and $(R11)$–$(R13)$, which regulate the propagation of disjunction in the presence of nested occurrences and make explicit the order-sensitive behaviour induced by McCarthy-type errors. These rules ensure that the deterministic calculus faithfully mirrors the non-commutative semantics of the deterministic five-valued matrix $\mathbf{M}^5_{BMK}$ introduced in the previous section.

Moreover, the sequent calculus $SC_{5}$ follows the structure of the deterministic systems in~\cite{Avron} closely, with modifications that reflect the semantics of the five-valued, non-commutative matrix introduced above. As in the original framework, the calculus consists of the standard structural axioms and negation rules, together with a family of introduction and propagation rules for the disjunction $\newor$, whose behaviour mirrors the deterministic but order-sensitive interpretation of the connective.	
The main difference concerns the right introduction rules $(R4_a)$ and $(R4_b)$, which generalize the single rule in~\cite{Avron} for disjunction into two distinct cases, as already done in the three-valued dynamic system.



\begin{table}[t]
\begin{mathpar}
	\inferrule*[left=A1]{}{\Gamma,\alpha \Rightarrow \Delta,\alpha} \quad
	\and 
	\inferrule*[left=A2]{}{\Gamma,\alpha,\neg\alpha \Rightarrow \Delta}
	\and
\end{mathpar}
\begin{mathpar}
	\inferrule*[left=r1]{\Gamma,\alpha \Rightarrow \Delta}{\Gamma,\neg\neg\alpha \Rightarrow \Delta}
	\and 
	\inferrule*[left=r2]{\Gamma \Rightarrow \Delta,\alpha}{\Gamma \Rightarrow \Delta,\neg\neg\alpha}
	\and
\end{mathpar}
\begin{mathpar}
	\inferrule*[left=r3]{\Gamma,\alpha \Rightarrow \Delta \quad \Gamma,\beta \Rightarrow \Delta}
	{\Gamma,\alpha\lor\beta \Rightarrow \Delta}
	\and
\end{mathpar}
\begin{mathpar}
	\inferrule*[left=r$4_a$]{\Gamma \Rightarrow \Delta,\alpha \quad \Gamma \Rightarrow \Delta,\neg\beta}{\Gamma \Rightarrow \Delta,\alpha\lor\beta}
	\and 
	\inferrule*[left=r$4_b$]{\Gamma \Rightarrow \Delta,\alpha \qquad \Gamma \Rightarrow \Delta,\beta}{\Gamma \Rightarrow \Delta,\alpha\lor\beta}
	\and 
	\inferrule*[left=r5]{\Gamma \Rightarrow \Delta,\neg\alpha \quad \Gamma \Rightarrow \Delta,\beta}{\Gamma \Rightarrow \Delta,\alpha\lor\beta}
\end{mathpar}
\begin{mathpar}
	\inferrule*[left=r$6_a$]{\Gamma\Rightarrow\Delta,\beta \quad \Gamma, \alpha \lor \beta \Rightarrow\Delta}{\Gamma, \neg \alpha\lor\beta \Rightarrow\Delta}
	\and 
	\inferrule*[left=r$7_a$]{\Gamma\Rightarrow\Delta, \beta \quad \Gamma\Rightarrow\Delta, \alpha \lor \beta }{\Gamma\Rightarrow\Delta,\neg \alpha\lor\beta}
	\and 
\end{mathpar}
\begin{mathpar}
	\inferrule*[left=r$6_b$]{\Gamma \Rightarrow \Delta, \alpha \qquad \Gamma, \alpha \lor \beta \Rightarrow \Delta}{\Gamma, \alpha \lor \neg \beta \Rightarrow \Delta}
	\and 
	\inferrule*[left=r$7_b$]{\Gamma \Rightarrow \Delta, \alpha \qquad \Gamma \Rightarrow \Delta, \alpha \lor \beta }{\Gamma \Rightarrow \Delta, \alpha \lor \neg \beta}
	\and 
\end{mathpar}
\begin{mathpar}
	\inferrule*[left=r8]{\Gamma,\neg\alpha \Rightarrow \Delta}{\Gamma,\neg(\alpha\lor\beta)\Rightarrow \Delta}
	\and 
	\inferrule*[left=r10]{\Gamma,\neg\beta \Rightarrow \Delta}{\Gamma,\neg(\alpha\lor\beta)\Rightarrow \Delta}
	\and 
	\inferrule*[left=r9]{\Gamma\Rightarrow\Delta,\neg\alpha \quad \Gamma\Rightarrow\Delta,\neg\beta}{\Gamma\Rightarrow\Delta,\neg(\alpha\lor\beta)}
\end{mathpar}
\begin{mathpar}
	\inferrule*[left=r11]{\Gamma \Rightarrow\Delta, \gamma \quad \Gamma, \alpha \lor \gamma \Rightarrow\Delta}{\Gamma,  (\alpha\lor\beta) \lor \gamma \Rightarrow\Delta}
	\and 
	\inferrule*[left=r12]{\Gamma\Rightarrow\Delta, \alpha \lor \gamma \quad \Gamma\Rightarrow\Delta, \beta \lor \gamma }{\Gamma\Rightarrow\Delta,(\alpha\lor\beta) \lor \gamma }
	\and 
	\inferrule*[left=r13]{\Gamma, \alpha \Rightarrow\Delta \quad \Gamma, \beta \lor \gamma \Rightarrow\Delta}{\Gamma, (\alpha\lor\beta) \lor \gamma \Rightarrow\Delta}
\end{mathpar}
\caption{Sequent calculus $SC_{5}$.}\label{tab:sc5}
\end{table}


In summary, the calculus $SC_{5}$ preserves the structural discipline of the original system of~\cite{Avron} while adapting it to our setting with three distinct error values.
The extension from four to five truth values allows the calculus to distinguish among soft (Kleene), medium (McCarthy), and critical (Bochvar) errors, each governed by a specific propagation pattern reflected in the rules above.

We now turn to the meta-theoretic analysis of the calculus $SC_{5}$. As in the three-valued case, soundness and completeness are established by relating the proof system to the deterministic semantics induced by the matrix $\mathbf{M}^{5}_{BMK}$.

Before proving soundness, we establish a basic algebraic property of the connective $\lor$ that will be used in the verification of some inference rules.

\begin{lemma}\label{lm:ssctvt}
The operation $\lor$ defined in the matrix $\mathbf{M}^{5}_{BMK}$ is associative.
\end{lemma}

 Making use of Lemma~\ref{lm:ssctvt} where necessary, soundness follows from a direct inspection of the axioms and inference rules, each of which is shown to preserve validity with respect to the five-valued semantics. The proof is routine and therefore omitted.

\begin{theorem}[Soundness of $SC_{5}$]\label{th:snd5}
The sequent calculus $SC_{5}$ is sound with respect to the deterministic five-valued semantics induced by the matrix $\mathbf{M}^{5}_{BMK}$.
\end{theorem}

Completeness is proved by means of a canonical countermodel construction. Given an unprovable sequent $\Gamma \Rightarrow \Delta$, one associates with
it a saturated unprovable pair $(T,S)$ extending $(\Gamma,\Delta)$. The construction of such pairs follows the same pattern adopted in the three-valued case and is therefore omitted here.

From a saturated pair $(T,S)$ one defines a canonical valuation $v:\mathcal{L}\to V$, assigning truth values in a hierarchical way so as to distinguish between Kleene-, McCarthy-, and Bochvar-type failures.

The canonical valuation $v:\mathcal{L}\to V$ associated with $(T,S)$ is defined by the following hierarchical clauses, ordered by decreasing severity of failure:

\[
v(\varphi)=
\begin{cases}
1 & \text{if } \varphi\in T,\\[4pt]
0 & \text{if } \neg\varphi\in T,\\[6pt]
k & \text{if } (\exists \beta\in T)\big(\varphi\lor\beta\in S\ \text{and}\ \beta\lor\varphi\in S\big),\\[6pt]
e & \text{if } (\exists \beta\in T)\big(\varphi\lor\beta\in S\big) \text{ and } (\forall \gamma\in T)\big(\varphi\lor\gamma\notin S\ \text{or}\ \gamma\lor\varphi\notin S\big),\\[6pt]
u & \text{otherwise.}
\end{cases}
\]
The idea is that the assignment captures the three degrees of failure:
\begin{itemize}
		
\item $u$ models \emph{Kleene uncertainty}---a soft, recoverable failure;
		
\item $e$ captures \emph{McCarthy sensitivity}---an order-dependent error, occurring when a right-directed propagation appears;
		
\item $k$ represents a \emph{Bochvar-type critical failure}, a symmetric blocking condition when both sides of a disjunction are rejected.
\end{itemize}
The definition of the valuation is shown to be well-formed, univocal, and semantically adequate: every formula receives exactly one value, and the assignment faithfully reflects the deterministic behaviour of the
five-valued disjunction. These properties are established by the following lemmas, culminating in the Truth Lemma.

\begin{lemma}[Mutual exclusivity]\label{lm:mtcl}
For every formula $\varphi$, exactly one of the clauses in the canonical definition of $v$ applies.
\end{lemma}

\begin{lemma}[Truth Lemma] \label{lm:trth}
\mbox{}
\begin{enumerate}
\item The implications “if” in points (i)--(iv) of the canonical clauses for $v$ can be replaced with equivalences ``iff''.
\item The ``iff'' version of those clauses holds for every formula $\alpha\in W$.
\end{enumerate}
\end{lemma}
\begin{theorem}[Completeness of $SC_{5}$]\label{th:cmpl5}
If a sequent $\Gamma\Rightarrow\Delta$ is valid in every valuation for the deterministic five-valued semantics of $SC_{5}$, then $\Gamma\Rightarrow\Delta$ is provable in $SC_{5}$.
\end{theorem}

\subsection{Relation between three-valued and five-valued semantics}
We now make precise in which sense the deterministic five-valued system can be understood as a formal refinement of the nondeterministic three-valued semantics introduced in Section~4. 
	
\begin{definition}
Let $\pi : \{0,1,u,e,k\} \to \{0,1,k\}$ be the projection function defined by:
\[
\pi(0)=0, \quad \pi(1)=1, \quad \pi(u)=\pi(e)=\pi(k)=k.
\]
We extend $\pi$ to valuations by setting $(\pi \circ v)(\varphi) := \pi(v(\varphi))$ for every formula $\varphi$.
\end{definition}
	
\begin{theorem}[Projection]\label{th:prjct}
Let $v$ be a deterministic valuation in $M^5_{BMK}$. Then $\pi \circ v$ is a dynamic valuation in the three-valued Nmatrix $M^3_{BMK}$.
\end{theorem}
	
The proof proceeds by structural induction on the formulas. The critical step lies in the evaluation of disjunction, which requires showing that the projection $\pi$ preserves the dynamic validity across the two semantics. Specifically, by mapping the values $\{u, e, k\}$ of $M^5_{BMK}$ into the single indeterminate value $k$ of $M^3_{BMK}$, it is straightforward to verify that the propagation of conflicts and errors in the five-valued matrix naturally complies with the nondeterministic absorbent operations of the three-valued Nmatrix.
\begin{remark}
This result establishes that the five-valued deterministic logic is an \emph{informational refinement} of the three-valued dynamic system. What appears as genuine non-determinism at the three-valued level (e.g., $1 \newor k = \{1,k\}$) is formally explained by $M^5_{BMK}$ as a deterministic choice hidden by the collapse of the error taxonomy. Therefore, the three-valued dynamic semantics acts as a sound abstraction for computing agents that lack the capacity to distinguish between specific error types.
\end{remark}

\section{Conclusions}\label{sc:conc}

In this paper, we have examined the logical treatment of nondeterministic choice in the presence of computational errors, arguing that standard three-valued logics fail to capture the symmetry required by genuinely nondeterministic choice. To overcome this limitation, we introduced a commutative and genuinely nondeterministic disjunction $\newor$ based on nondeterministic matrices, combining Kleene and Bochvar behaviours within a unified semantic framework.
We provided sound and complete sequent calculi for both dynamic and static interpretations of the resulting semantics, clarifying different modes of nondeterminism resolution. We further showed how this framework can be deterministically refined into a five-valued system accounting for graded forms of computational failure and sequential behaviour. These results support the view that nondeterministic matrices constitute a natural semantic setting for reasoning about symmetric choice in the presence of error.

The study of our calculi plays an essential role in the setting of process algebraic approaches to concurrency theory, as they provide a formal bridge between logical specifications and the temporal behavior of concurrent systems. This synergy offers significant advantages for formal verification: specifically, the \emph{compositionality} of the sequent calculus is directly inherited by the process algebra. This ensures that verification tasks can be handled modularly: error-free components do not require re-verification when integrated into a larger system, as their logical properties are preserved by the structural rules of the calculus. Consequently, valid sequent proofs serve as formal certificates, guaranteeing that the corresponding processes satisfy critical safety properties regarding error propagation and nondeterministic choice resolution.

Several directions for further investigation arise from the present study. A central one concerns the development of an explicit equational theory for the sequent calculi introduced here, including the identification of algebraic counterparts of the dynamic and static systems, their equational bases. Since the semantics considered involves nondeterministic matrices, such an investigation cannot rely solely on standard universal algebra, but naturally calls for more general frameworks, such as partial algebras or nondeterministic algebras (see~\cite{Burmeister86, Carnielli16, Reichel22}).
This perspective is promising both for relating the connective $\newor$ to algebraic models of computation, such as process algebras, and for clarifying its connections with well-known three-valued varieties. More generally, an algebraic investigation of the five-valued deterministic semantics—possibly in terms of non-commutative algebraic structures—may clarify how intrinsically nondeterministic matrices, such as $\mathbf{M}^3_{BMK}$ and $\mathbf{M}^3_{MK}$~\cite{Avron}, can be embedded into a deterministic setting.

Finally, the formalisation of the new connective $\newor$ paves the way for defining an extension of the CCS-like process algebra~\cite{AM25} mentioned in Section~\ref{sc:3vld}. In particular, the properties of $\newor$ can be leveraged to define a complete axiomatisation for the process algebra, which would rely on Eq.~(\ref{axiom:newor}). Moreover, the sequent calculi we have proposed can provide the base for a formal proof system to verify compositionally properties of algebraic process terms expressed in the version of the Hennessy-Milner logic used in~\cite{AM25}, similarly as done, e.g., in~\cite{SIMPSON2004287}.

\nocite{*}
\bibliographystyle{eptcs}
\bibliography{biblio}
\clearpage

\end{document}